\documentclass[preprint,11pt]{elsarticle}

\usepackage{amssymb}
\usepackage{xspace}

\newcommand{\pions}{$\pi^+\pi^-\pi^0$\xspace}
\newcommand{\piz}{$\pi^0$\xspace}
\newcommand{\pip}{$\pi^+$\xspace}
\newcommand{\pim}{$\pi^-$\xspace}

\newcommand{\jpsi}{$J/\psi$\xspace}
\newcommand{\psip}{$\psi^\prime$\xspace}

\newcommand{\jpsidecay}{$J/\psi \longrightarrow \pi^+\pi^-\pi^0$\xspace}
\newcommand{\psipdecay}{$\psi^\prime \longrightarrow \pi^+\pi^-\pi^0$\xspace}
\newcommand{\jpsidecaymath}{J/\psi \longrightarrow \pi^+\pi^-\pi^0}
\newcommand{\psipdecaymath}{\psi^\prime \longrightarrow \pi^+\pi^-\pi^0}

\newcommand{\ccbar}{$c\bar{c}$\xspace}

\newcommand{\jpsibfresultmath}{(\resultJpsiPermilleBranchingFraction \pm \resultJpsiPermilleBranchingFractionStatError~(stat.) ^{+ \resultJpsiPermilleBranchingFractionSysErrorUp}_{- \resultJpsiPermilleBranchingFractionSysErrorDown}~(syst.) ^{+ \resultJpsiPermilleBranchingFractionNormErrorUp}_{- \resultJpsiPermilleBranchingFractionNormErrorDown}~(norm.)) \times 10^{-2}}

\newcommand{\psibfresultmath}{(\resultPsiPermilleBranchingFraction \pm \resultPsiPermilleBranchingFractionStatError~(stat.) ^{+ \resultPsiPermilleBranchingFractionSysErrorUp}_{- \resultPsiPermilleBranchingFractionSysErrorDown}~(syst.) ^{+ \resultPsiPermilleBranchingFractionNormErrorUp}_{- \resultPsiPermilleBranchingFractionNormErrorDown}~(norm.)) \times 10^{-4}}

\newcommand{\resultJpsiPermilleBranchingFraction}{2.137}
\newcommand{\resultJpsiPermilleBranchingFractionStatError}{0.004}
\newcommand{\resultJpsiPermilleBranchingFractionSysErrorUp}{0.058}
\newcommand{\resultJpsiPermilleBranchingFractionSysErrorDown}{0.056}
\newcommand{\resultJpsiPermilleBranchingFractionNormErrorUp}{0.027}
\newcommand{\resultJpsiPermilleBranchingFractionNormErrorDown}{0.026}

\newcommand{\resultJpsiTotdatamillion}{225.2}
\newcommand{\resultJpsiNdata}{1859771}
\newcommand{\resultJpsiErrorNdata}{1364}
\newcommand{\resultJpsiNcont}{8811}
\newcommand{\resultJpsiErrorNcont}{1582}
\newcommand{\resultJpsiNincl}{9919}
\newcommand{\resultJpsiErrorNincl}{463}
\newcommand{\resultJpsiMCEfficiency}{38.66}
\newcommand{\resultJpsiErrorMCEfficiency}{0.05}

\newcommand{\resultPsiPermilleBranchingFraction}{2.14}
\newcommand{\resultPsiPermilleBranchingFractionStatError}{0.03}
\newcommand{\resultPsiPermilleBranchingFractionSysErrorUp}{0.08}
\newcommand{\resultPsiPermilleBranchingFractionSysErrorDown}{0.07}
\newcommand{\resultPsiPermilleBranchingFractionNormErrorUp}{0.09}
\newcommand{\resultPsiPermilleBranchingFractionNormErrorDown}{0.08}

\newcommand{\resultPsiTotdatamillion}{106.4}
\newcommand{\resultPsiNdata}{7872}
\newcommand{\resultPsiErrorNdata}{89}
\newcommand{\resultPsiNcont}{820}
\newcommand{\resultPsiErrorNcont}{55}
\newcommand{\resultPsiNincl}{101}
\newcommand{\resultPsiErrorNincl}{32}
\newcommand{\resultPsiMCEfficiency}{30.91}
\newcommand{\resultPsiErrorMCEfficiency}{0.14}

\begin{document}

\begin{frontmatter}

\title{Precision measurement of the branching fractions of $J/\psi \rightarrow \pi^+\pi^-\pi^0$ and $\psi^\prime \rightarrow \pi^+\pi^-\pi^0$}
\author{
\begin{small}
M.~Ablikim$^{1}$, M.~N.~Achasov$^{5}$, D.~Alberto$^{41}$, D.J.~Ambrose$^{38}$, F.~F.~An$^{1}$, Q.~An$^{39}$, Z.~H.~An$^{1}$, J.~Z.~Bai$^{1}$, R.~B.~F.~Baldini Ferroli$^{17}$, Y.~Ban$^{25}$
% ...
\\
\vspace{0.2cm}
(BESIII Collaboration)\\
\vspace{0.2cm} {\it
$^{1}$ Institute of High Energy Physics, Beijing 100049, P. R. China\\
% ...
$^{5}$ G.I. Budker Institute of Nuclear Physics SB RAS (BINP), \\Novosibirsk 630090, Russia\\
% ...
$^{17}$ INFN Laboratori Nazionali di Frascati , Frascati, Italy\\
% ...
$^{25}$ Peking University, Beijing 100871, P. R. China\\
% ...
$^{38}$ University of Rochester, Rochester, New York 14627, USA\\
$^{39}$ University of Science and Technology of China, Hefei 230026, P. R. China\\
% ...
$^{41}$ University of Turin and INFN, Turin, Italy\\
% ...
\vspace{7cm}
}
\end{small}
}

\begin{abstract}
We study the decays of the \jpsi and \psip mesons to \pions using data 
samples at both resonances collected with the BES III detector in 2009.
We measure the corresponding branching fractions with unprecedented 
precision and provide mass spectra and Dalitz plots. The branching 
fraction for \jpsidecay is determined to be 
	\[
\jpsibfresultmath,
\]
and the branching fraction for \psipdecay is measured as
	\[
\psibfresultmath.
\]
The \jpsi decay is found to be dominated by an intermediate $\rho(770)$ state, whereas the \psip decay is dominated by di-pion masses around 2.2~GeV/$c^2$, leading to strikingly different Dalitz distributions.
\end{abstract}


\end{frontmatter}

\section{Introduction}
\label{Introduction}

Previous studies of \jpsidecay
\cite{Franklin:1983ve,Bai:2004jn,Aubert:2004kj,Aubert:2007ef} and 
\psipdecay \cite{Franklin:1983ve, Adam:2004pr, Ablikim:2005jy} found 
not only an unexpectedly low branching fraction in the case of the 
\psip\footnote{Some authors \cite{Brodsky:1997fj} claim that the 
\psipdecay branching fraction is as expected and the \jpsidecay 
fraction is much higher than expected for a \ccbar model of the 
\jpsi.} (world averages: $BF(\jpsidecaymath) = (2.07 \pm 0.12) \times 
10^{-2}$ and $BF(\psipdecaymath) = (1.68 \pm 0.26)\times 10^{-4}$ 
\cite{PDG2010}) but also a completely different shape of the di-pion 
mass spectrum and the Dalitz plot. The fact that the 
$\rho(770)\pi$ decays as a fraction of all hadronic decays are 
suppressed by two orders of magnitude in the \psip with regards to the 
\jpsi is especially difficult to explain and known as the \emph{$\rho \pi$ 
puzzle}. Suggested solutions include intrinsic charm in the light 
vector mesons \cite{Brodsky:1997fj}, formation of three-gluon 
intermediate resonances \cite{Brodsky:1987bb}, a hybrid nature of the 
\psip \cite{Kisslinger:2008xn}, an additional hadronic amplitude for 
the \psip decays \cite{Suzuki:2001fs}, the \jpsi being dominantly a 
higher Fock-state \cite{Chen:1998ma} and so on.

In this letter we present new measurements of the \jpsidecay and 
\psipdecay branching fractions with unprecedented precision using the 
large data samples collected with the BES III detector at the \jpsi and \psip 
resonances. These measurements are an important first step to an 
experimental inquiry into the puzzle of the decay dynamics,
preparing the way for a detailed analysis. The large branching 
fraction of \jpsidecay also makes it an important background process 
for many other studies (e.g. the scalar meson spectrum in $J/\psi 
\rightarrow \gamma \pi^+\pi^-$), an improved knowledge of this branching 
fraction will thus also enhance the precision of those measurements.

\section{Detector and Monte Carlo Simulation}
\label{Detector}

BEPC~II is a double-ring $e^{+}e^{-}$ collider designed to provide a
peak luminosity of $10^{33}$ cm$^{-2}s^{-1}$ at a beam current of
$0.93$~A. The BES~III \cite{Ablikim:2009vd} detector has a geometrical acceptance of
$93\%$ of $4\pi$ and has four main components: (1) A small-cell,
helium-based ($40\%$ He, $60\%$ C$_{3}$H$_{8}$) main drift chamber
(MDC) with $43$ layers providing an average single-hit resolution of
$135$~$\mu$m, charged-particle momentum resolution in a $1$~T
magnetic field of $0.5\%$ at 1~GeV$/c$, and a $dE/dx$ resolution
that is better than $6\%$. (2) An electromagnetic calorimeter (EMC)
consisting of $6240$ CsI(Tl) crystals in a cylindrical structure
(barrel) and two end caps. The energy resolution at $1.0$~GeV$/c$ is
$2.5\%$ ($5\%$) in the barrel (end caps), and the position resolution
is $6$~mm ($9$~mm) in the barrel (end caps). (3) A time-of-flight
system (TOF) constructed of $5$-cm-thick plastic scintillators, with
$176$ detectors of $2.4$~m length in two layers in the barrel and
$96$ fan-shaped detectors in the end caps. The barrel (end cap) time
resolution of $80$~ps ($110$~ps) provides a $2\sigma$ $K/\pi$
separation for momenta up to $\sim 1.0$~GeV$/c$. (4) The muon system
(MUC) consists of $1000$~m$^{2}$ of Resistive Plate Chambers (RPCs)
in nine barrel and eight end cap layers and provides $2$~cm position
resolution.

For the events to be read-out, one out of seven trigger conditions 
based on combinations of signals from the MDC, TOF and EMC had to be 
fulfilled. At least one of these rather loose conditions should always 
be fulfilled for the events under study, and indeed overall trigger 
efficiencies very close to 100\% were found for hadronic events 
containing charged particles \cite{Berger2010}.

The efficiencies of the detector and the event selection are estimated 
using a Monte Carlo (MC) simulation based on \textsc{Geant4} 
\cite{Agostinelli:2002hh, Allison:2006ve}. \textsc{evtgen} 
\cite{Ping2008} is used to generate events; for the \jpsidecay decay, 
$\rho(770)\pi$ events give a good description of the data, while for 
\psipdecay a mixture of $\rho(770)\pi$ and $P$-wave phase-space 
events is used \footnote{The use of $P$-wave phase space (and 
subsequent reweighting in the Dalitz-variables) is motivated by the 
angular distributions and the fact that not much is known about the
dynamics leading to the accumulation of events around a di-pion mass 
of 2.2~GeV/$c^2$. This procedure has been checked using both toy MC 
samples and a sample generated using the amplitudes extracted from 
a phenomenological fit to the data. The maximum difference in 
efficiency obtained is taken as a systematic error.}. In both cases, 
differences between the generated and observed distributions are taken 
care of by reweighting the MC events to the data distribution in the 
Dalitz variables. For the estimation of backgrounds, inclusive MC 
samples are generated by \textsc{kkmc} 
\cite{Jadach:1999vf,Jadach:2000ir} --- known decays of the \jpsi and 
\psip are modeled by \textsc{evtgen} according to the branching 
fractions provided by the Particle Data Group (PDG) \cite{PDG2010}, 
and the remaining unknown decay modes are generated with 
\textsc{Lundcharm} \cite{Ping2008}. Backgrounds from the process 
$J/\psi \rightarrow \gamma \pi^+\pi^-$ have been modeled using 
amplitudes extracted from a partial wave analysis of BES~III data.

\section{Data Samples and Event Selection}
\label{Selection}

\begin{figure}[tb!]
	\centering
		\includegraphics[width=0.45\textwidth]{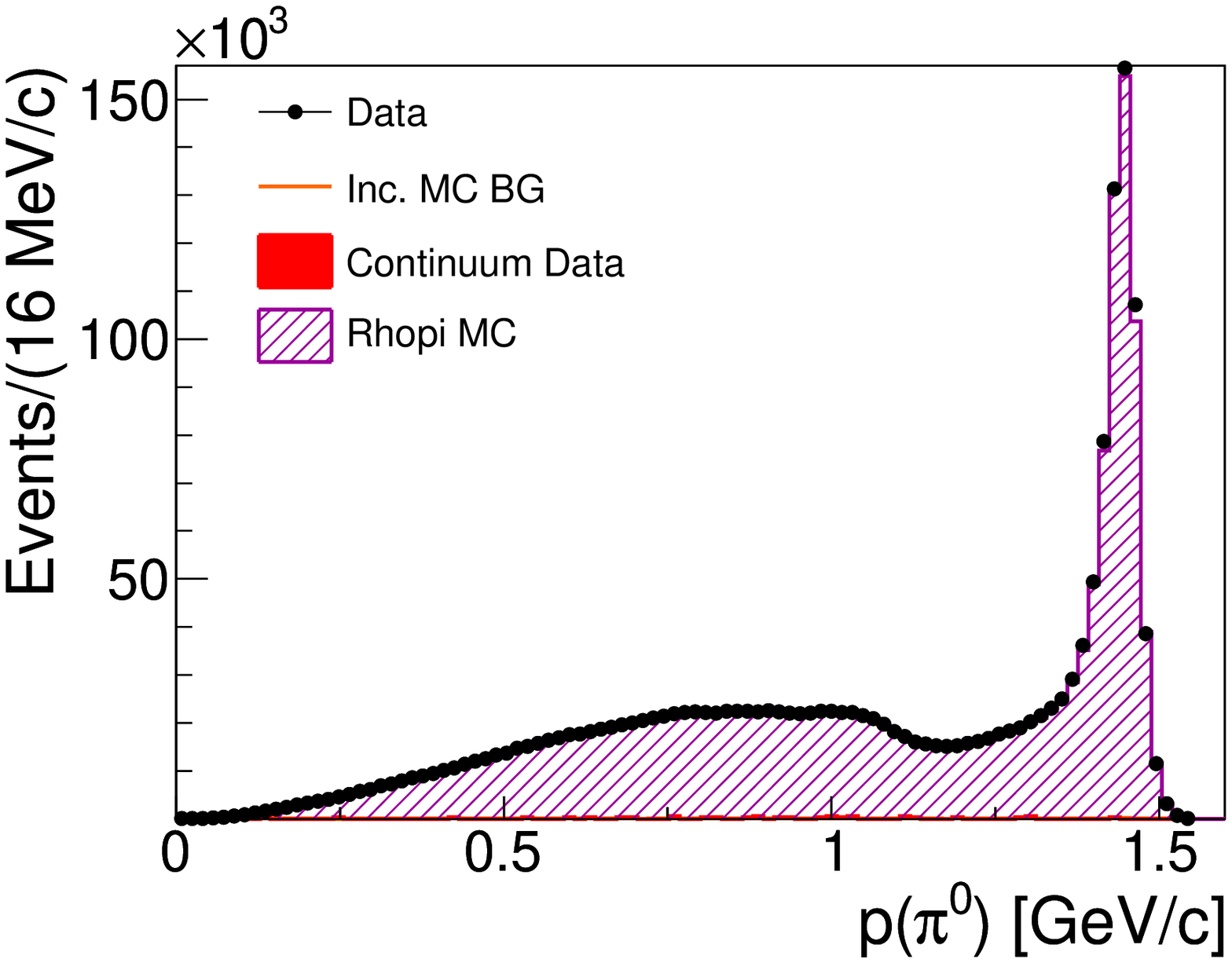}
		\includegraphics[width=0.45\textwidth]{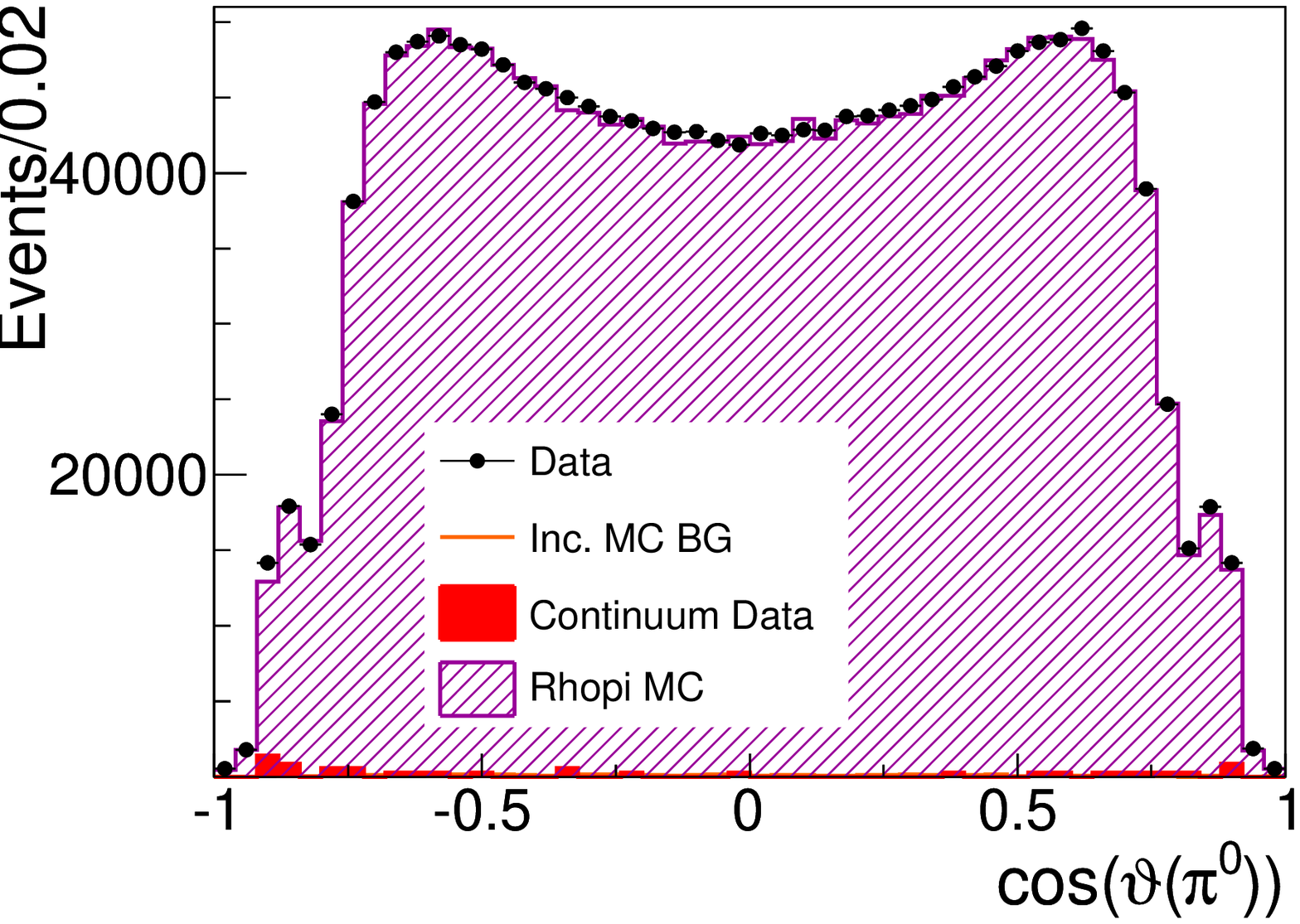}
		\includegraphics[width=0.45\textwidth]{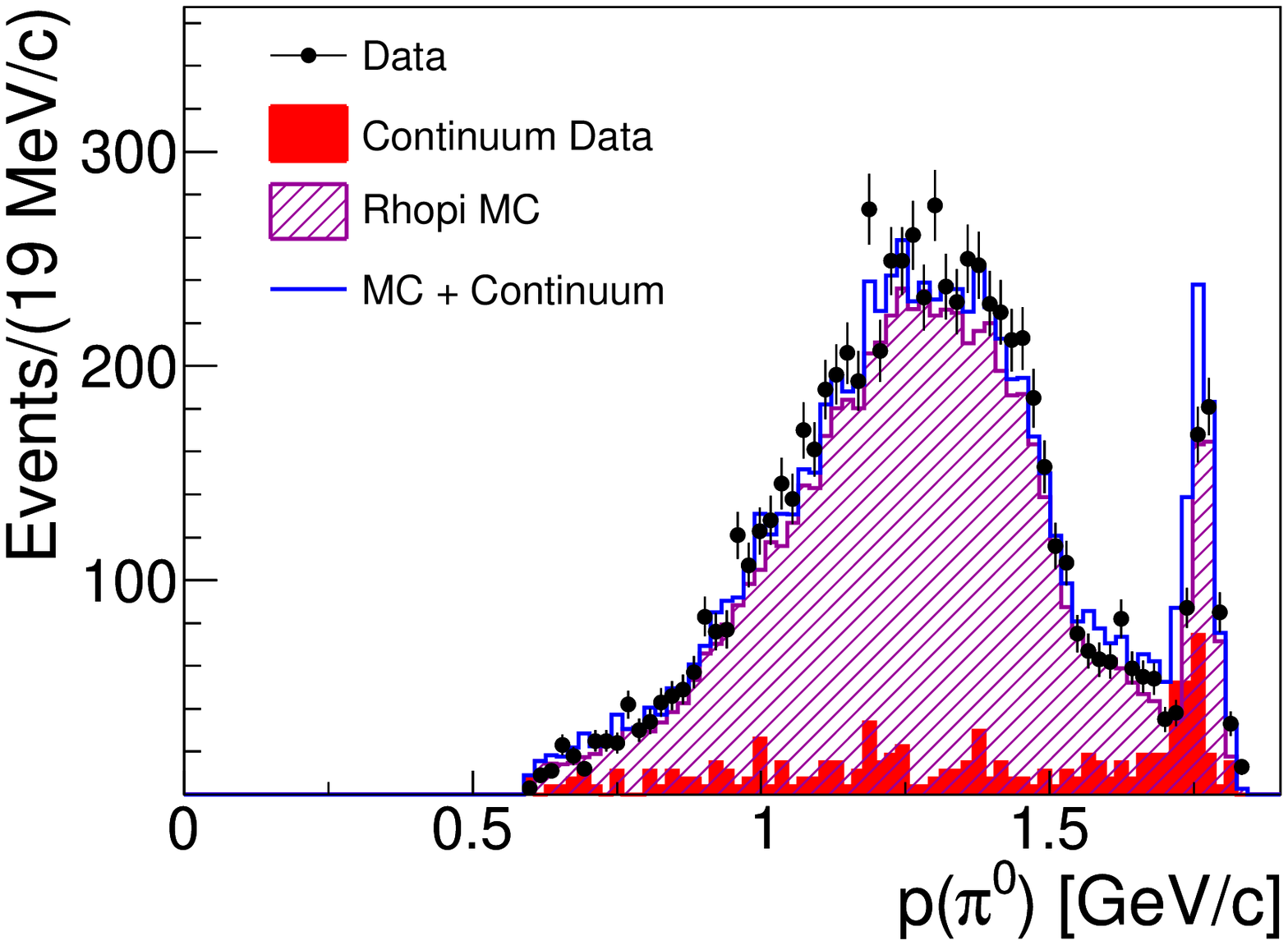}
		\includegraphics[width=0.45\textwidth]{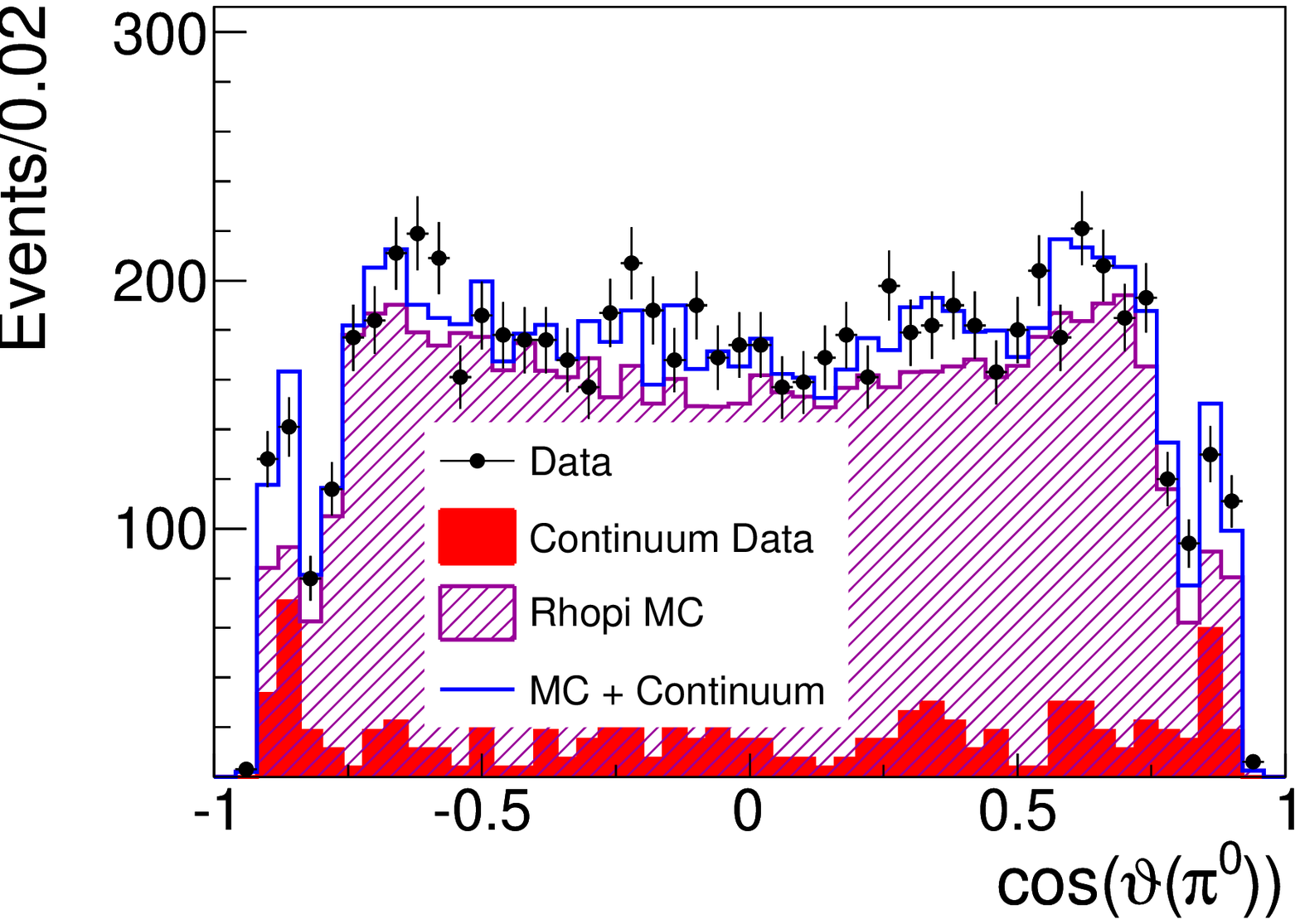}
	\caption{Kinematical distributions of the reconstructed $\pi^0$s: Top for the 
	  \jpsidecay analysis, bottom for the \psipdecay analysis; left 
	  showing the \piz momentum, right showing the \piz polar 
	  angle distributions.}
	\label{fig:pi0kinematics}
\end{figure}

This analysis uses a sample of $2.25 \times 10^{8}$ $J/\psi$
decays~\cite{Ablikim:2010kp} and $1.06 \times 10^{8}$ $\psi^{\prime}$
decays~\cite{Ablikim:2010zn} collected by BES~III in 2009. 

Charged particle tracks in BES~III are reconstructed using MDC 
hits. We require tracks to pass within $\pm 10$ cm from the 
interaction point in the beam direction, within 1~cm of the beam line 
in the plane perpendicular to the beam and to have a polar angle in 
the range $|\cos \vartheta| < 0.93$. Events are required to contain 
exactly one track of positive and one of negative charge, 
corresponding to the \pip and \pim. Electromagnetic showers are 
reconstructed by clustering EMC crystal energies \cite{He2011}. The energy deposit 
in nearby TOF counters is included in order to improve the 
reconstruction efficiency and energy resolution. Showers identified as 
photon candidates must satisfy fiducial, timing and shower-quality 
requirements. Showers from the barrel region ($|\cos \vartheta| < 
0.8$) are required to have an energy above 25~MeV, while those in the 
end caps ($0.86 < |\cos \vartheta| < 0.92$) must have at least 
50~MeV. Showers from the transition region between barrel and end cap 
are excluded from the analysis, as are showers within 10$^\circ$ from 
any charged track. Events are required to contain at least two showers 
fulfilling these criteria.

For every pair of photon candidates, a full event kinematic fit with 
the initial particle (\jpsi or \psip) four-momentum as a constraint is 
performed. The pair leading to the smallest $\chi^2$ is kept as the 
\piz candidate if $\chi^2 < 50$, otherwise, the event is rejected. The 
fit is repeated once more with the assumption that the charged 
particles are kaons; if this leads to a smaller $\chi^2$, the event is 
also rejected. Yet another kinematic fit is performed with the mass of 
the \piz as an additional constraint; the resulting $\chi^2$ is 
required to be less than 50. The invariant mass of the two photon 
candidates has to be compatible with the mass of the \piz, 
$0.11$~GeV/$c^2$ $< m_{\gamma\gamma} <$~0.15~GeV/$c^2$.

For the \psipdecay analysis, additional requirements are needed to 
suppress backgrounds from radiative decays to $e^+e^-$, $\mu^+\mu^-$ 
and the $J/\psi$ and $\chi_c$ states, namely the invariant mass of the 
charged pion candidates is required to be less than 3~GeV/$c^2$, the 
energy deposits associated to the tracks is required to be less than 
0.8~GeV and the penetration depth into the muon system less than 
40~cm.

\section{Efficiency correction}
\label{effcor}

In order to study differences between the simulation and data in track
reconstruction, an analysis of a specially selected \jpsi to $3 \pi$
candidate subsample with one or two tracks and two or more photons and
with tight requirements on one track and the \piz is
performed. Specifically, the TOF and $\frac{dE}{dx}$ information of
the charged track are combined to form particle identification
confidence levels for the $\pi$ and $K$ hypotheses; the likelihood for
the $\pi$ hypothesis is then required to be larger than the likelihood
for the $K$ hypothesis.  Kinematic fits to the \piz mass are performed
for all pairs of photon candidates, and the pair with the lowest
$\chi^2$ is taken as the \piz candidate, if the $\chi^2$ is less than
20. The invariant mass of the object recoiling against the system of
the track and the \piz is required to be between 0 and 0.2~GeV$/c^2$,
and the recoil direction must be within the tracker acceptance.  Using
these tagged events, the efficiency for finding and correctly
reconstructing the second track is determined. The simulation is then
corrected as a function of polar angle and track momentum to reflect
the efficiency found in data (which is on average about 2\% lower than
the simulated efficiency).

Similarly, an analysis using a subsample similar to the one above but
requiring two tracks and with tight requirements on the two tracks and
photons, chosen with the standard photon selection, is performed to test
the \piz reconstruction efficiency. The tracks are again required to
pass particle identification requirements as above and in addition are
required to have an associated energy deposit in the calorimeter of
less than 0.8 times the beam energy to remove electrons and a
penetration depth in the muon system less than 40~cm to remove
muons. Their opening angle is required to be less than
170$^\circ$. The momentum of the system recoiling against the tracks
must be larger than 100~MeV$/c$, and the invariant mass of the tracks
and the two photon candidates must be above 3.0~GeV$/c^2$.  Here the
efficiency differences found between selected data and simulated
events (of the order of 0.5\%) are used to correct the simulation as a
function of \piz momentum. Figure \ref{fig:pi0kinematics} shows the
reconstructed kinematics of the \piz for the selected events compared
to the corrected MC simulation.

\section{Results}
\label{Results}

\begin{table}
	\centering
		\begin{tabular}{lrr}
		\hline
		Quantity & \jpsidecay & \psipdecay \\
		\hline
		$N_{sel}$ 						& \resultJpsiNdata~$\pm$ \resultJpsiErrorNdata & \resultPsiNdata~$\pm$ \resultPsiErrorNdata\\
		$N^{BG}_{continuum}$ 	& \resultJpsiNcont~$\pm$ \resultJpsiErrorNcont & \resultPsiNcont~$\pm$ \resultPsiErrorNcont\\
		$N^{BG}_{resonance}$ 	& \resultJpsiNincl~$\pm$ \resultJpsiErrorNincl & \resultPsiNincl~$\pm$ \resultPsiErrorNincl\\
		$N_{\psi}$					  & (\resultJpsiTotdatamillion~$\pm$ 2.8)~Million & (\resultPsiTotdatamillion~$\pm$ 4)~Million \\
		$\epsilon_{MC}$ 			& \resultJpsiMCEfficiency~$\pm$~\resultJpsiErrorMCEfficiency & 		
																															\resultPsiMCEfficiency~$\pm$~\resultPsiErrorMCEfficiency\\
		$\epsilon_{trig}$ 		& \multicolumn{2}{c}{\hspace{5mm} (100 - 0.2)\% \hspace{3mm} \cite{Berger2010}} \\
		$BF(\pi^0 \rightarrow \gamma\gamma)$ & \multicolumn{2}{c}{\hspace{5mm} (98.823 $\pm$ 0.023)\% \hspace{3mm}  \cite{PDG2010}} \\
		\hline	
		\\
		\end{tabular}
	\caption{Numbers used in the branching fraction calculation}
	\label{tab:NumbersUsedInTheBranchingFractionCalculation}
\end{table}

1,859,771 events from the \jpsi sample and 7872 events from the \psip 
sample survive all selection criteria. The branching fractions are 
calculated as follows:
\begin{equation}
	BF = \frac{N_{sel} - N^{BG}_{continuum} - N^{BG}_{resonance}}{N_\psi \cdot \epsilon_{MC} \cdot \epsilon_{trig} \cdot BF(\pi^0 \rightarrow \gamma \gamma)},
\end{equation}
where $N_{sel}$ is the number of selected events, $N^{BG}_{continuum}$ 
the number of background events from the continuum (estimated from 
data samples taken at 3.08~GeV and 3.65~GeV), $N^{BG}_{resonance}$ the 
number of background events from other resonance processes (estimated 
using inclusive MC samples) and $N_\psi$ the number of \jpsi 
or \psip mesons in the sample. $\epsilon_{MC}$ is the 
efficiency determined from signal MC, 
$\epsilon_{trig}$ is the trigger efficiency (found to be very close to 100\%, \cite{Berger2010}), 
and the branching fraction for $\pi^0 \rightarrow \gamma \gamma$ is taken 
from the PDG \cite{PDG2010} --- the corresponding numbers can be found in table \ref{tab:NumbersUsedInTheBranchingFractionCalculation}. In this calculation, interference of resonance and continuum processes is neglected.

\begin{table}
	\centering
\begin{tabular}{lrrrr}
\hline 
& \multicolumn{2}{c}{\jpsidecay} & \multicolumn{2}{c}{\psipdecay}\\
\hline
       & Upward & Downward & Upward & Downward\\ 
Source of systematic & Change & Change   & Change & Change\\ 
 & (\%) & (\%) & (\%) & (\%) \\ 
\hline 
MC simulation&0.25 &-0.23 & 1.20 &-1.20\\ 
EMC Energy scale&0.02 & -0.02 & 0.18 &-0.15\\ 
$\gamma$ efficiency&2.04 &-1.96 & 2.04 &-1.96\\ 
$\pi^{0}$ kinematic fit& 0.28 &-0.27 & 0.27 &-0.27\\ 
tracking efficiency&1.64 &-1.59 & 1.80 &-1.75\\ 
Muon cut&--- & --- & 1.28 &-0.75\\ 
Trigger efficiency&0.20 &0.00 & 0.20 &0.00\\ 
Resonance background&0.67 &-0.67 & 1.45 &-1.45\\ 
\hline
Syst. w/o normalization &2.74& -2.64 & 3.57 & 3.33\\
Normalization&1.26 &-1.23 & 4.17 &-3.85\\ 
\hline
Total syst.~uncertainty& 3.01 &-2.91 & 5.49 &5.09\\
Syst. + stat.~uncertainty& 3.02 &-2.91 & 5.72 &5.34\\
\hline 
\end{tabular}
	\caption{Impact of the systematic uncertainties 
	  on the measured branching fractions; the various 
	  sources of systematic uncertainties lead to 
	  the listed upward and downward changes in the branching fractions.}
	\label{tab:sys}
\end{table}

\begin{figure}[tb!]
	\centering
		\includegraphics[width=0.45\textwidth]{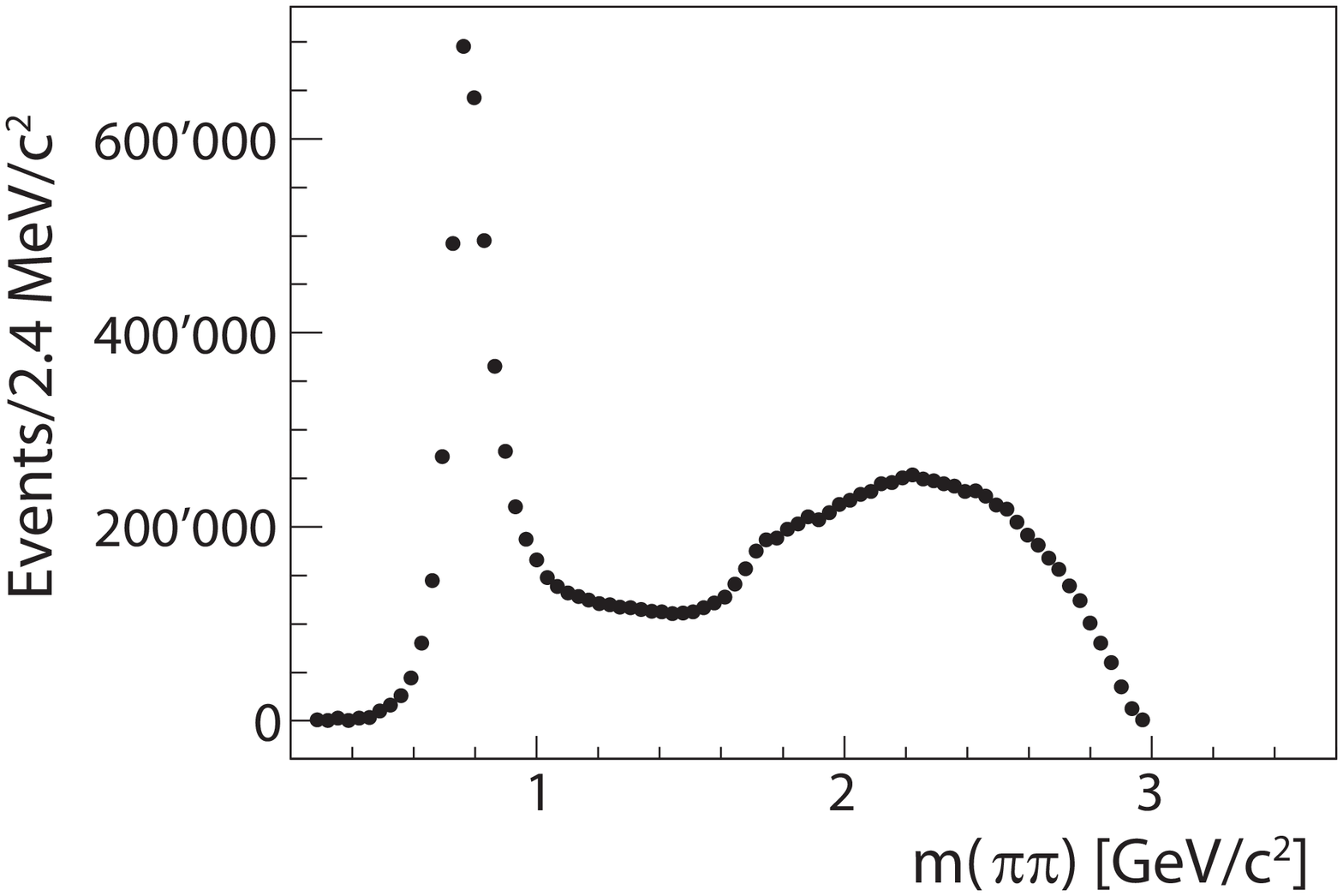}
		\includegraphics[width=0.45\textwidth]{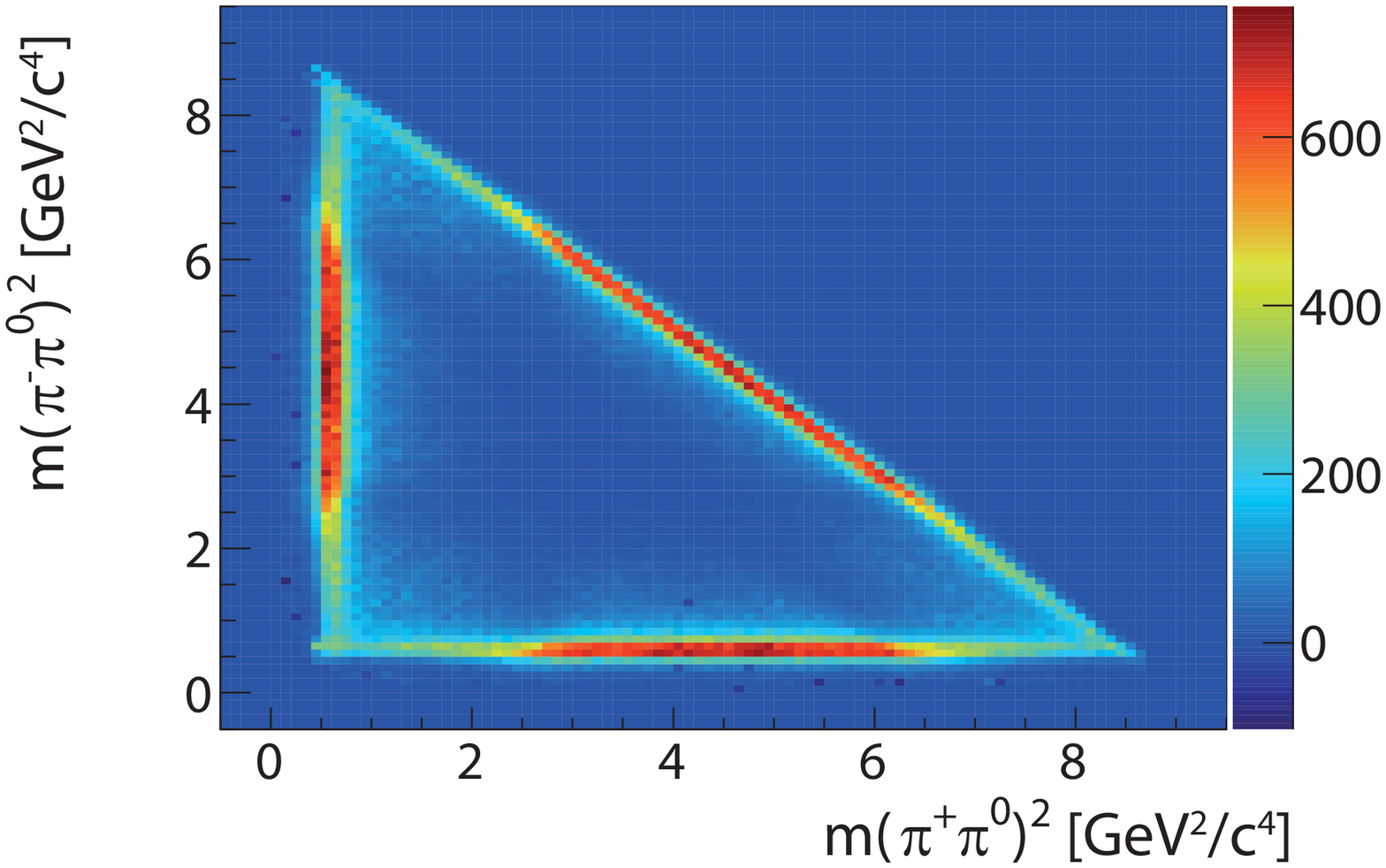}
		\includegraphics[width=0.45\textwidth]{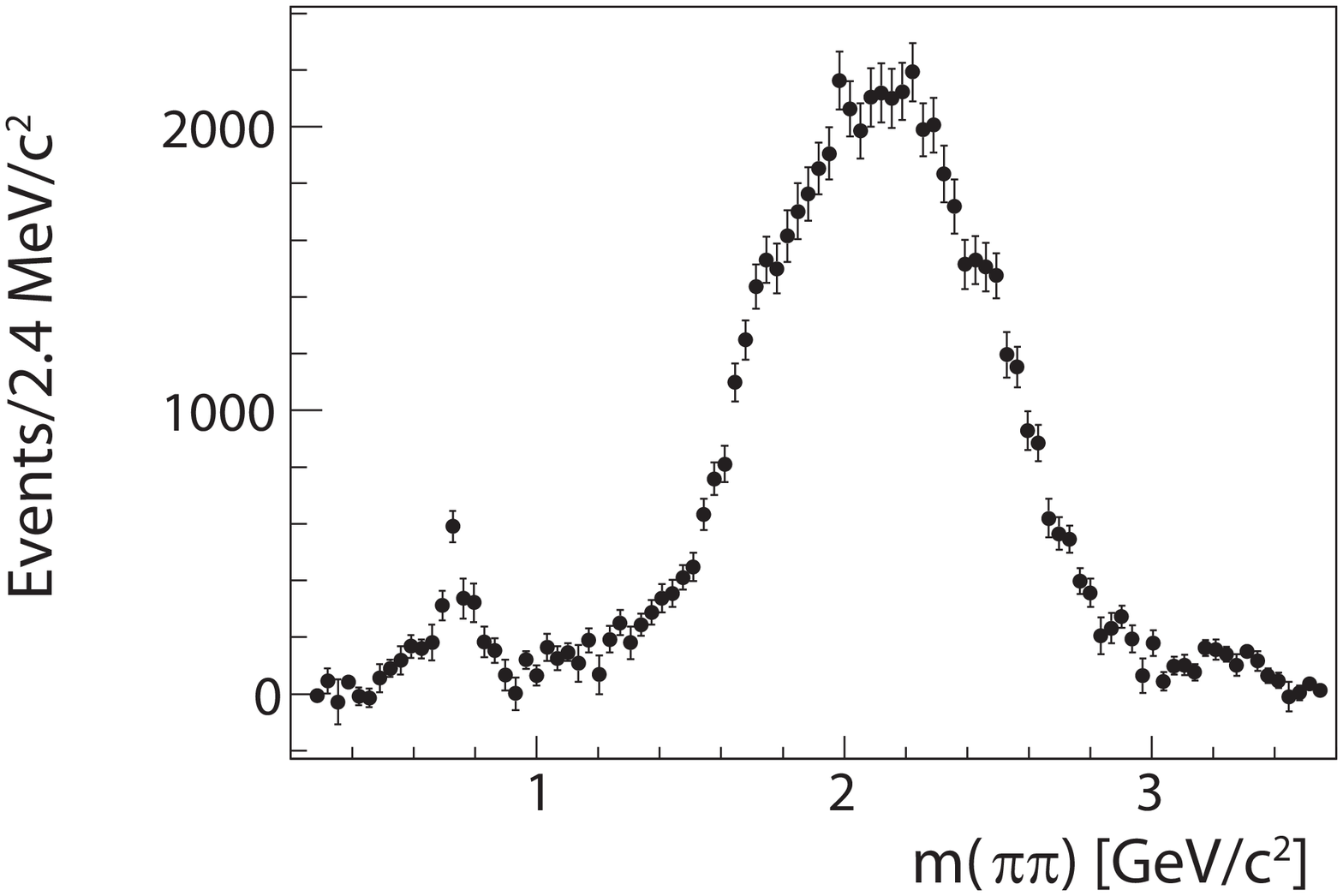}
		\includegraphics[width=0.45\textwidth]{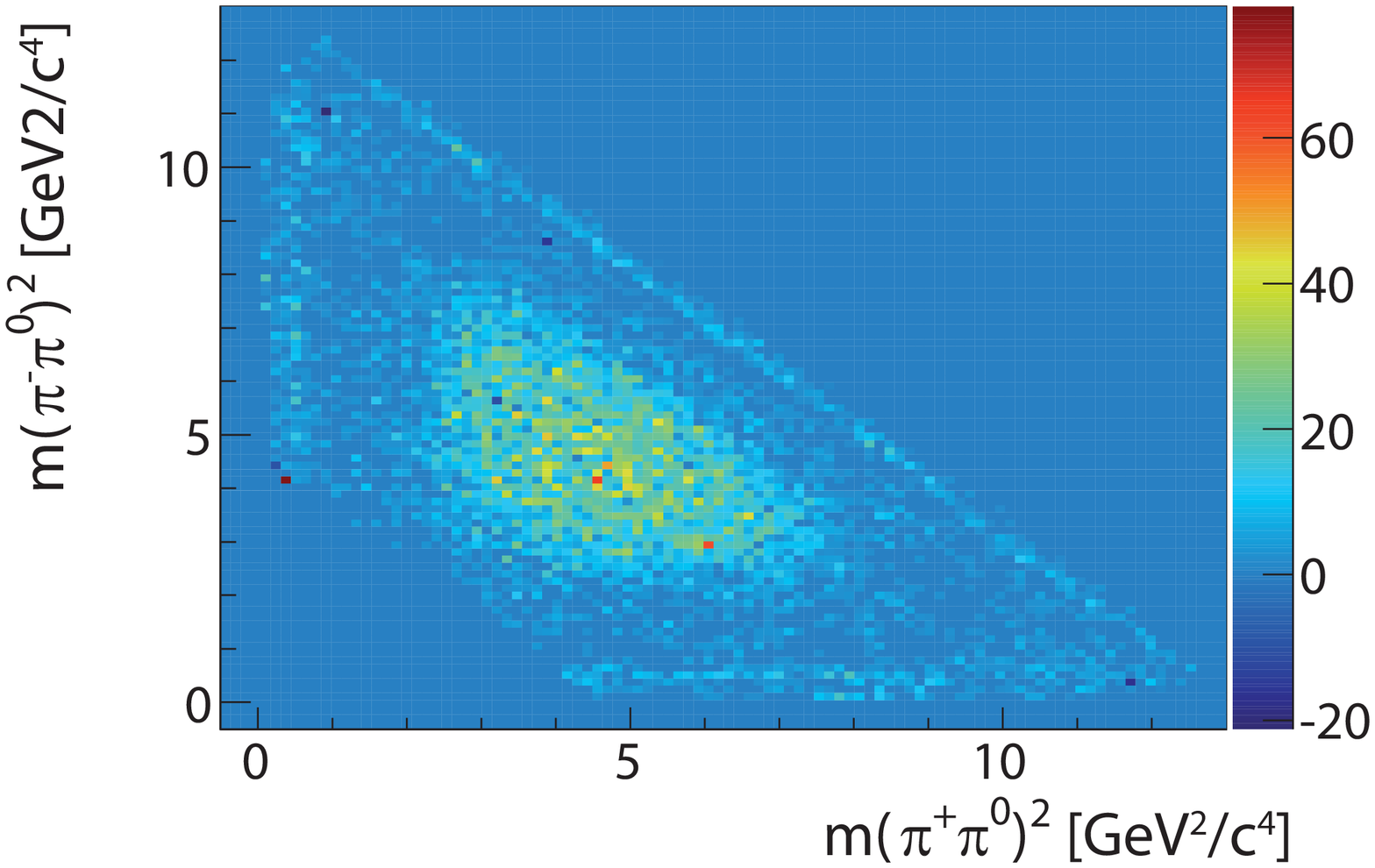}
	\caption{$\pi\pi$ invariant mass distribution (left) and Dalitz plot (right)  with backgrounds subtracted and corrected for efficiency. Top and bottom graphs show the results for the \jpsidecay and \psipdecay analysis, respectively.}
	\label{fig:results}
\end{figure}

Possible systematic errors resulting from the following sources were studied:
\begin{itemize}
\item The uncertainty due to the simulation model was estimated by the 
difference in the efficiency with and without the reweighting 
described in section \ref{Detector} for the \jpsi sample and by 
comparing with the efficiency obtained from a sample generated using 
amplitudes extracted from a phenomenological fit\footnote{In this fit, 
contributions from $\rho(770)$, a higher $\rho$ with a mass of 
2285~MeV/$c^2$ and a width of 950~MeV/$c^2$ and a $\rho^3$ with a mass of 1750~MeV/$c^2$ 
and a width of 650~MeV/$c^2$ were found to lead to an adequate description 
of the data.} for the \psip sample. For both cases, the model error is 
not the dominant systematic error.
\item The absolute energy scale of the EMC is known to an accuracy of 
  0.4\% \cite{Ablikim:2010rc};
\item The photon detection and reconstruction efficiency is described 
by the simulation to within 1\% per photon \cite{Ablikim:2010rc}.
\item The uncertainty due to the \piz finding and kinematic fitting 
was estimated by performing a different analysis with the \jpsi data 
sample (see section \ref{effcor}). A tighter selection was applied to the
charged tracks and no \piz was required. The difference between data 
and simulation is taken as the systematic error.
\item The uncertainty due to charged particle track finding and 
kinematic fitting was estimated using an analysis with tight 
requirements on the \piz and one charged track. The efficiencies for
finding and reconstructing the other track were compared between 
data and simulation (see section \ref{effcor}).
\item The efficiency of the muon rejection (used only in the \psip 
analysis) was estimated by either dropping the requirement or 
demanding a penetration less than $30$~cm instead of less than 
$40$~cm.
\item The trigger efficiency was changed from 100\% to 99.8\%, 
reflecting the statistical uncertainty of the efficiency determination 
\cite{Berger2010}.
\item The background from continuum processes was estimated using 
samples taken off-resonance (282~nb$^{-1}$ of luminosity taken at a 
center-of-mass energy of 3.08 GeV, compared to 81~pb$^{-1}$ at the 
\jpsi resonance and 43~pb$^{-1}$ taken at a center-of-mass energy of 
3.650~GeV, compared to 163~pb$^{-1}$ at the \psip resonance). The 
small samples due to the clean selection lead to relatively large 
statistical errors for the continuum contribution (18.0\% for the 
\jpsi and 6.7\% for the \psip). Compared to these errors the 
systematic errors from the luminosity measurements or varying 
beam conditions can be neglected.
\item The accuracy of the inclusive simulation for describing 
background from resonant processes was checked in analyses requiring 
one photon less (a \pip\pim$\gamma$ final state) or one photon more 
(a \pip\pim\piz$\gamma$ final state) and was found to be mediocre; 
it is assigned an uncertainty of 100\%.
\item The normalization (number of $J/\psi$ or $\psi'$ events) has an
uncertainty of 1.23\% for the \jpsi sample \cite{Ablikim:2010kp} and
4\% for the \psip sample \cite{Ablikim:2010zn}.
\end{itemize}
Table \ref{tab:sys} shows the impact of the systematic errors on the measured branching fractions.

The branching fraction for \jpsidecay is determined to be 
	\[
\jpsibfresultmath,
\]
and the branching fraction for \psipdecay is measured as
	\[
\psibfresultmath.
\]

Invariant mass spectra and Dalitz plots are shown in figure 
\ref{fig:results}. The decay \jpsidecay is dominated by $\rho(770)$ 
production; the absence of events in the center of the Dalitz plot 
points to negatively interfering higher $\rho$ states. In the case of 
the \psipdecay decay, a small $\rho(770)$ contribution can be 
discerned. Most of the events are however clustering around 
2.2~GeV$/c^2$ in di-pion mass. To disentangle the contributions of 
various excited $\rho$ states to this peak will require a partial wave 
analysis.

\section{Conclusion}
\label{Conclusion}
The branching fractions for \jpsidecay and \psipdecay have been 
measured with unprecedented precision at the BES III experiment. The 
measurement for \jpsidecay is in good agreement with the world average 
of $BF(\jpsidecaymath) = (2.07\pm0.12)\times10^{-2}$ \cite{PDG2010} 
while the result for \psipdecay is slightly larger than the world 
average of $BF(\psipdecaymath) = (1.68 \pm 0.26) \times 10^{-4}$ 
\cite{PDG2010}. The ratio of these two branching fractions
	\[
\frac{BF(\psipdecaymath)}{BF(\jpsidecaymath)} = (1.00 \pm 0.01~(stat.) ^{+0.06}_{-0.05}~(syst.))\%,
\]
where correlations between the systematic errors of the two analyses have been taken into account,
is found to be an order of magnitude smaller than the ratio of $12$\% naively 
expected from the fraction of decays via three gluon exchange. 

The 
decay dynamics of the \jpsi are dominated by the $\rho(770)$ 
meson. While the $\rho(770)$ is also visible in the case of the \psip 
decay, the dynamics there is dominated by states at higher 
masses. Understanding the nature of these higher mass states and why 
they are suppressed in \jpsi decays and enhanced in \psip decays may 
be clarified in a partial wave analysis, which is beyond the scope of 
this letter.

\section{Acknowledgement}
\label{sec:Acknowledgement}

The BESIII collaboration thanks the staff of BEPCII and the computing center for their hard efforts. This work is supported in part by the Ministry of Science and Technology of China under Contract No. 2009CB825200; National Natural Science Foundation of China (NSFC) under Contracts Nos. 10625524, 10821063, 10825524, 10835001, 10935007; Joint Funds of the National Natural Science Foundation of China under Contract No. 11079008; the Chinese Academy of Sciences (CAS) Large-Scale Scientific Facility Program; CAS under Contracts Nos. KJCX2-YW-N29, KJCX2-YW-N45; 100 Talents Program of CAS; Istituto Nazionale di Fisica Nucleare, Italy; Siberian Branch of Russian Academy of Science, joint project No 32 with CAS; U. S. Department of Energy under Contracts Nos. DE-FG02-04ER41291, DE-FG02-91ER40682, DE-FG02-94ER40823; U.S. National Science Foundation; University of Groningen (RuG) and the Helmholtzzentrum fuer Schwerionenforschung GmbH (GSI), Darmstadt; WCU Program of National Research Foundation of Korea under Contract No. R32-2008-000-10155-0; Swiss National Science Foundation.

\bibliographystyle{model1-num-names}
\bibliography{literature}

\begin{thebibliography}{23}
\expandafter\ifx\csname natexlab\endcsname\relax\def\natexlab#1{#1}\fi
\providecommand{\bibinfo}[2]{#2}
\ifx\xfnm\relax \def\xfnm[#1]{\unskip,\space#1}\fi
\bibitem[{Franklin et~al.(1983)}]{Franklin:1983ve}
\bibinfo{author}{M.~E.~B. Franklin}, et~al.,
\newblock \bibinfo{title}{Measurement of psi (3097) and psi-prime (3686) decays
  into selected hadronic modes},
\newblock \bibinfo{journal}{Phys. Rev. Lett.} \bibinfo{volume}{51}
  (\bibinfo{year}{1983}) \bibinfo{pages}{963}.
\bibitem[{Bai et~al.(2004)}]{Bai:2004jn}
\bibinfo{author}{J.~Z. Bai}, et~al.,
\newblock \bibinfo{title}{{Measurement of the Branching Fraction of $J/\psi
  \longrightarrow \pi^+ \pi^- \pi^0$}},
\newblock \bibinfo{journal}{Phys. Rev.} \bibinfo{volume}{D70}
  (\bibinfo{year}{2004}) \bibinfo{pages}{012005}.
\bibitem[{Aubert et~al.(2004)}]{Aubert:2004kj}
\bibinfo{author}{B.~Aubert}, et~al.,
\newblock \bibinfo{title}{{Study of $e^+e^-$ to $\pi^+ \pi^- \pi^0$ process
  using initial state radiation with BaBar}},
\newblock \bibinfo{journal}{Phys. Rev.} \bibinfo{volume}{D70}
  (\bibinfo{year}{2004}) \bibinfo{pages}{072004}.
\bibitem[{Aubert et~al.(2007)}]{Aubert:2007ef}
\bibinfo{author}{B.~Aubert}, et~al.,
\newblock \bibinfo{title}{{The $e^+ e^-$ to $2(\pi^+\pi^-)\pi^0$,
  $2(\pi^+\pi^-)\eta$, $K^+ K^-\pi^+\pi^-\pi^0$ and $K^+ K^-\pi^+\pi^-\eta$
  Cross Sections Measured with Initial-State Radiation}},
\newblock \bibinfo{journal}{Phys. Rev.} \bibinfo{volume}{D76}
  (\bibinfo{year}{2007}) \bibinfo{pages}{092005}.
\bibitem[{Adam et~al.(2005)}]{Adam:2004pr}
\bibinfo{author}{N.~E. Adam}, et~al.,
\newblock \bibinfo{title}{{Observation of $1^-(0^-)$ final states from
  $\psi(2S)$ decays and $e^+e^-$ annihilation}},
\newblock \bibinfo{journal}{Phys. Rev. Lett.} \bibinfo{volume}{94}
  (\bibinfo{year}{2005}) \bibinfo{pages}{012005}.
\bibitem[{Ablikim et~al.(2005)}]{Ablikim:2005jy}
\bibinfo{author}{M.~Ablikim}, et~al.,
\newblock \bibinfo{title}{{Partial wave analysis of $\psi^\prime \to \pi^+
  \pi^- \pi^0$ at BESII}},
\newblock \bibinfo{journal}{Phys. Lett.} \bibinfo{volume}{B619}
  (\bibinfo{year}{2005}) \bibinfo{pages}{247--254}.
\bibitem[{Brodsky and Karliner(1997)}]{Brodsky:1997fj}
\bibinfo{author}{S.~Brodsky}, \bibinfo{author}{M.~Karliner},
\newblock \bibinfo{title}{{Intrinsic charm of vector mesons: A possible
  solution of the \emph{rho pi puzzle}}},
\newblock \bibinfo{journal}{Phys. Rev. Lett.} \bibinfo{volume}{78}
  (\bibinfo{year}{1997}) \bibinfo{pages}{4682--4685}.
\bibitem[{Nakamura et~al.(2010)}]{PDG2010}
\bibinfo{author}{K.~Nakamura}, et~al.,
\newblock \bibinfo{title}{{Review of particle physics}},
\newblock \bibinfo{journal}{J.Phys.G} \bibinfo{volume}{G37}
  (\bibinfo{year}{2010}) \bibinfo{pages}{075021}.
\bibitem[{Brodsky et~al.(1987)Brodsky, Lepage, and Tuan}]{Brodsky:1987bb}
\bibinfo{author}{S.~Brodsky}, \bibinfo{author}{G.~P. Lepage},
  \bibinfo{author}{S.~F. Tuan},
\newblock \bibinfo{title}{{Exclusive charmonium decays: The $J/\psi$
  ($\psi^\prime$) $\to \rho \pi, K^* \bar{K}$ puzzle}},
\newblock \bibinfo{journal}{Phys. Rev. Lett.} \bibinfo{volume}{59}
  (\bibinfo{year}{1987}) \bibinfo{pages}{621}.
\bibitem[{Kisslinger et~al.(2009)Kisslinger, Parno, and
  Riordan}]{Kisslinger:2008xn}
\bibinfo{author}{L.~Kisslinger}, \bibinfo{author}{D.~Parno},
  \bibinfo{author}{S.~Riordan},
\newblock \bibinfo{title}{{Hybrid Charmonium and the $\rho-\pi$ Puzzle}},
\newblock \bibinfo{journal}{Adv. High Energy Phys.} \bibinfo{volume}{2009}
  (\bibinfo{year}{2009}) \bibinfo{pages}{982341}.
\bibitem[{Suzuki(2001)}]{Suzuki:2001fs}
\bibinfo{author}{M.~Suzuki},
\newblock \bibinfo{title}{{Possible hadronic excess in psi (2S) decay and the
  rho pi puzzle}},
\newblock \bibinfo{journal}{Phys. Rev.} \bibinfo{volume}{D63}
  (\bibinfo{year}{2001}) \bibinfo{pages}{054021}.
\bibitem[{Chen and Braaten(1998)}]{Chen:1998ma}
\bibinfo{author}{Y.~Chen}, \bibinfo{author}{E.~Braaten},
\newblock \bibinfo{title}{{An Explanation for the rho-pi Puzzle of J/psi and
  psi' Decays}},
\newblock \bibinfo{journal}{Phys. Rev. Lett.} \bibinfo{volume}{80}
  (\bibinfo{year}{1998}) \bibinfo{pages}{5060--5063}.
\bibitem[{Ablikim et~al.(2010)}]{Ablikim:2009vd}
\bibinfo{author}{M.~Ablikim}, et~al.,
\newblock \bibinfo{title}{{Design and Construction of the BESIII Detector}},
\newblock \bibinfo{journal}{Nucl.Instrum.Meth.} \bibinfo{volume}{A614}
  (\bibinfo{year}{2010}) \bibinfo{pages}{345--399}.
\bibitem[{Berger et~al.(2010)}]{Berger2010}
\bibinfo{author}{N.~Berger}, et~al.,
\newblock \bibinfo{title}{{Trigger efficiencies at BES III}},
\newblock \bibinfo{journal}{Chin. Phys.} \bibinfo{volume}{C34}
  (\bibinfo{year}{2010}) \bibinfo{pages}{1779}.
\bibitem[{Agostinelli et~al.(2003)}]{Agostinelli:2002hh}
\bibinfo{author}{S.~Agostinelli}, et~al.,
\newblock \bibinfo{title}{{GEANT4: A Simulation toolkit}},
\newblock \bibinfo{journal}{Nucl.Instrum.Meth.} \bibinfo{volume}{A506}
  (\bibinfo{year}{2003}) \bibinfo{pages}{250--303}.
\bibitem[{Allison et~al.(2006)Allison, Amako, Apostolakis, Araujo, Dubois
  et~al.}]{Allison:2006ve}
\bibinfo{author}{J.~Allison}, \bibinfo{author}{K.~Amako},
  \bibinfo{author}{J.~Apostolakis}, \bibinfo{author}{H.~Araujo},
  \bibinfo{author}{P.~Dubois}, et~al.,
\newblock \bibinfo{title}{{Geant4 developments and applications}},
\newblock \bibinfo{journal}{IEEE Trans.Nucl.Sci.} \bibinfo{volume}{53}
  (\bibinfo{year}{2006}) \bibinfo{pages}{270}.
\bibitem[{Ping(2008)}]{Ping2008}
\bibinfo{author}{R.~Ping},
\newblock \bibinfo{title}{{Event generators at BES III}},
\newblock \bibinfo{journal}{Chin. Phys.} \bibinfo{volume}{C 32}
  (\bibinfo{year}{2008}) \bibinfo{pages}{599}.
\bibitem[{Jadach et~al.(2000)Jadach, Ward, and Was}]{Jadach:1999vf}
\bibinfo{author}{S.~Jadach}, \bibinfo{author}{B.~Ward},
  \bibinfo{author}{Z.~Was},
\newblock \bibinfo{title}{{The Precision Monte Carlo event generator K K for
  two fermion final states in e+ e- collisions}},
\newblock \bibinfo{journal}{Comput.Phys.Commun.} \bibinfo{volume}{130}
  (\bibinfo{year}{2000}) \bibinfo{pages}{260--325}.
\bibitem[{Jadach et~al.(2001)Jadach, Ward, and Was}]{Jadach:2000ir}
\bibinfo{author}{S.~Jadach}, \bibinfo{author}{B.~Ward},
  \bibinfo{author}{Z.~Was},
\newblock \bibinfo{title}{{Coherent exclusive exponentiation for precision
  Monte Carlo calculations}},
\newblock \bibinfo{journal}{Phys.Rev.} \bibinfo{volume}{D63}
  (\bibinfo{year}{2001}) \bibinfo{pages}{113009}.
\bibitem[{Ablikim et~al.(2011)}]{Ablikim:2010kp}
\bibinfo{author}{M.~Ablikim}, et~al.,
\newblock \bibinfo{title}{{Measurement of the Matrix Element for the Decay
  $\eta^{\prime} \to \eta \pi^+\pi^-$}},
\newblock \bibinfo{journal}{Phys. Rev.} \bibinfo{volume}{D83}
  (\bibinfo{year}{2011}) \bibinfo{pages}{012003}.
\bibitem[{Ablikim et~al.(2010)}]{Ablikim:2010zn}
\bibinfo{author}{M.~Ablikim}, et~al.,
\newblock \bibinfo{title}{{Branching fraction measurements of $\chi_{c0}$ and
  $\chi_{c2}$ to $\pi^0\pi^0$ and $\eta\eta$}},
\newblock \bibinfo{journal}{Phys. Rev.} \bibinfo{volume}{D81}
  (\bibinfo{year}{2010}) \bibinfo{pages}{052005}.
\bibitem[{He(2011)}]{He2011}
\bibinfo{author}{M.~He},
\newblock \bibinfo{title}{{Simulation and reconstruction of the BESIII EMC}},
\newblock \bibinfo{journal}{Journal of Physics: Conference Series}
  \bibinfo{volume}{293} (\bibinfo{year}{2011}) \bibinfo{pages}{012025}.
\bibitem[{Ablikim et~al.(2010)}]{Ablikim:2010rc}
\bibinfo{author}{M.~Ablikim}, et~al.,
\newblock \bibinfo{title}{{Measurements of $h_c(^1P_1)$ in $\psi^\prime$
  Decays}},
\newblock \bibinfo{journal}{Phys. Rev. Lett.} \bibinfo{volume}{104}
  (\bibinfo{year}{2010}) \bibinfo{pages}{132002}.

\end{thebibliography}

\end{document}